\documentclass[aip,reprint,amsmath,amssymb]{revtex4-1}

\usepackage{graphicx}
\usepackage{dcolumn}
\usepackage{float}
\usepackage{bm}
\usepackage{amssymb}
\usepackage{amsmath}
\usepackage{lipsum}

\newcommand{\be}{\begin{eqnarray}}
\newcommand{\ee}{\end{eqnarray}}
\newcommand{\cP}{\mbox{$\cal{P} $}}
\newcommand{\nn}{\nonumber}
\newcommand{\bes}{\begin{eqnarray*}}
\newcommand{\ees}{\end{eqnarray*}}

\begin{document}

\title{Free Energy of a Knotted Polymer Confined to Narrow Cylindrical and Conical Channels}

\author{James M. Polson}
%\email{jpolson@upei.ca}
\affiliation{Department of Physics, University of Prince Edward Island, 
550 University Ave., Charlottetown, Prince Edward Island, C1A 4P3, Canada}
\author{Cameron Hastie}
\altaffiliation{Current address: Department of Physics, McGill University, 3600 rue University, 
Montreal, Quebec, H3A 2T8, Canada}
\affiliation{Department of Physics, University of Prince Edward Island,
550 University Ave., Charlottetown, Prince Edward Island, C1A 4P3, Canada}

\date{\today}

\begin{abstract}
Monte Carlo simulations are used to study the conformational behavior of a semiflexible 
polymer confined to cylindrical and conical channels. The channels are sufficiently
narrow that the conditions for the Odijk regime are marginally satisfied. For cylindrical 
confinement, we examine polymers with a single knot of topology $3_1$, $4_1$, or $5_1$, as 
well as unknotted polymers that are capable of forming S-loops. We measure the variation of 
the free energy $F$ with the end-to-end polymer extension length $X$ and examine the effect 
of varying the polymer topology, persistence length $P$ and cylinder diameter $D$ on the 
free energy functions. Similarly, we characterize the behavior of the knot span along 
the channel. We find that increasing the knot complexity increases the typical size of 
the knot. In the regime of low $X$, where the knot/S-loop size is large, the conformational 
behavior is independent of polymer topology. In addition, the scaling properties of the 
free energy and knot span are in agreement with predictions from a theoretical model
constructed using known properties of interacting polymers in the Odijk regime.
We also examine the variation of $F$ with position of a 
knot in conical channels for various values of the cone angle $\alpha$. The free energy 
decreases as the knot moves in a direction where the cone widens, and it also decreases 
with increasing $\alpha$ and with increasing knot complexity. The behavior is in agreement
with predictions from a theoretical model in which the dominant contribution to the 
change in $F$ is the change in the size of the hairpins as the knot moves to the wider
region of the channel.
\end{abstract}

\maketitle

\section{Introduction}
\label{sec:intro}

In recent years, numerous experimental studies have contributed to the systematic investigation 
of the physical behavior of single DNA molecules confined to nanochannels.\cite{dai2016polymer,%
reisner2012dna} Enabled by advances in nanofabrication techniques, such work is 
mainly motivated by a variety of applications that exploit the effects of confinement on 
polymers, including DNA sorting,\cite{dorfman2012beyond} DNA denaturation mapping,%
\cite{reisner2010single,marie2013integrated} and genome mapping.\cite{lam2012genome,%
hastie2013rapid,dorfman2013fluid,muller2017optical} The development of nanofluidic devices 
for such purposes naturally benefits from a deep understanding of the conformational statistics 
and dynamics of polymers in nanochannels. 

One aspect of DNA that has received considerable attention in recent years is its propensity 
to form knots.\cite{orlandini2018statics} Knots can occur in DNA as byproducts of various
biological processes, including replication, transcription and recombination.%
\cite{orlandini2018statics} Confinement of DNA can dramatically increase the probability
and complexity of knot formation, as observed for example in knots in DNA extracted from 
viruses.\cite{arsuaga2005dna} Knots in DNA stretched by elongational fields or confinement in
channels have been observed and their 
dynamics characterized in a variety of {\it in vitro} experiments,\cite{rybenkov1993probability,%
bao2003behavior,ercolini2007fractal,tang2011compression,renner2015stretching,plesa2016direct,%
klotz2017dynamics,soh2018knots,amin2018nanofluidic,sharma2019complex,klotz2018motion,%
soh2018untying,metzler2006diffusion,reifenberger2015topological,amin2018nanofluidic,ma2020diffusion} 
where they are typically detected by the presence of bright spots in optical images of stained DNA 
molecules.  Knots can be created by a variety of methods, including tying individual DNA molecules 
using optical tweezers\cite{bao2003behavior} or by application of a strong alternating electric 
field in microfluidics devices that employ elongational fields.\cite{tang2011compression} 

Of particular relevance to the present work are those experiments which have examined knotted 
DNA confined to nanochannels.\cite{metzler2006diffusion,reifenberger2015topological,amin2018nanofluidic,%
ma2020diffusion} Understanding such systems is important for the development of next-generation 
genomics technology that use nanochannel mapping assays, where the presence of knots or backfolds 
introduces artifacts that may lead to erroneous results.\cite{reifenberger2015topological}
Several years ago, Reifenberger {\it et al.} examined ``topological events'' occurring along 
DNA molecules driven into narrow square channels (40--50 nm wide) using an analysis of spikes 
in the YOYO intensity profile.\cite{reifenberger2015topological} The presence of either backfolds 
or knots was evident from intensity spikes of approximately 3$\times$ that of the adjacent 
region.  It was noted that frequency of these structures was significantly less and their size
significantly greater than the values predicted from simulations.\cite{micheletti2014knotting,%
jain2017simulations} In another recent study, Amin {\it et al.} developed a nanofluidic ``knot 
factory'' which utilizes hydrodynamic compression of single DNA molecules against a barrier in 
wide (325~nm$\times$414~nm) rectangular nanochannels to form sequences of simple knots that 
can be studied upon subsequent extension of the molecule.\cite{amin2018nanofluidic}
They found that the knotting probability increases with chain compression and with waiting
time in the compressed state. They also noted a breakdown of Poisson statistics of knotting
probability at high compression, likely due to interactions between knots. Very recently,
Ma and Dorfman used the technique of Ref.~\onlinecite{amin2018nanofluidic} for knot generation
to study diffusion of knots in nanochannels in the extended de~Gennes regime.\cite{ma2020diffusion} 
They observed a subdiffusive motion, contradicting the prevailing theory for diffusion of knots 
in channel-confined DNA, but consistent with earlier observations of self-reptation of knots
for unconfined DNA under tension.  Note that the mechanisms used to create knots in such
experiments do not allow for selection of knots of a specific type, nor is the knot topology 
easy to characterize afterward, other than an approximate measurement of the DNA contour 
length contained in the knot. Indeed, the difficulty in distinguishing a knot from a simple 
backfold was noted in Ref.~\onlinecite{reifenberger2015topological}.

Numerous computer simulation studies of knotted polymers have contributed to 
elucidating the behavior of knots in DNA.\cite{micheletti2011polymers,orlandini2018statics} 
A number of these studies have examined the statics and dynamics of knotted polymers confined
to narrow channels.\cite{mobius2008spontaneous,micheletti2012knotting,orlandini2013knotting,%
nakajima2013localization,micheletti2014knotting,suma2015knotting,dai2015metastable,jain2017simulations}
M\"obius {\it et al.} examined the dynamics of a trefoil knot in a channel-confined semiflexible 
chain in the Odijk regime using Brownian dynamics (BD) simulations together with a coarse-grained 
theoretical model.\cite{mobius2008spontaneous} They concluded that the knot inflates to macroscopic size 
before untying.  As noted elsewhere,\cite{nakajima2013localization} however, their theoretical 
model omits the key feature of excluded-volume interactions between overlapping subchains, which 
tend to reduce knot size and may lead to knot localization instead of knot expansion. In subsequent
Monte Carlo simulation studies, such knot localization of a trefoil knot was observed for 
both flexible\cite{nakajima2013localization,dai2015origin} and semiflexible\cite{dai2015metastable} chains
under confinement in channels.
In the case of flexible chains, the typical knot size was observed to decrease monotonically with 
decreasing channel size, and the overall behavior was explained using a model employing the de~Gennes 
blob scaling and repulsion between blobs in different overlapping subchains in the confined 
knot.\cite{nakajima2013localization} For the case of confined semiflexible chains, the typical
knot size varied non-monotonically with decreasing channel width, initially increasing,
then reaching a maximum before decreasing as the channel becomes narrower.\cite{dai2015metastable}
This behavior was explained using a theoretical model for the free energy based on
an earlier theory of knots in unconfined wormlike polymers.\cite{grosberg2007metastable,%
dai2014metastable, dai2015origin} Other MC simulation studies have shown that the knotting
probability for DNA (with persistence length $\approx 50$~nm) peaks for channel widths
slightly below 100~nm, and that simpler knots (especially trefoil) tend to dominate.%
\cite{micheletti2012knotting,orlandini2013knotting,jain2017simulations} Langevin dynamics simulations 
suggest that the abrupt decrease in the knotting probability for channel widths below 
100~nm arises from a decrease in the knot lifetime and an increase in the mean time between the 
formation of knots for decreasing channel widths in this range.\cite{micheletti2014knotting,%
suma2015knotting}

In the present study we use MC simulations to examine the behavior of a single knotted
semiflexible polymer confined to a narrow channel in the Odijk regime. This regime
is defined by the condition that $P\gg D$, where $P$ is the persistence length and $D$
is the channel width, though the condition is only marginally satisfied in this work. 
With the exception of Ref.~\onlinecite{mobius2008spontaneous}, each of the simulation 
studies discussed above considered channels that correspond either to the extended de~Gennes 
scaling regime ($P\ll D \ll P^2/w$, where $w$ is the polymer width) or 
else the onset of Odijk scaling for channel widths near 50~nm.  Although the equilibrium
probabilities of knots in this regime are expected to be very low, it is nevertheless
of interest to extend the range of confinement over which knotting behavior is well
characterized, as noted in the conclusions of Ref.~\onlinecite{amin2018nanofluidic}. 
In addition, it is convenient for testing theoretical models in such a clearly defined
scaling regime. As various studies have noted that simple knots are most probable under
channel confinement, we choose to consider only knots of these types. We employ simulation 
methods similar to those used previously to study polymer folding in nanochannels%
\cite{polson2017free,polson2018free} and measure the variation in the free energy with
respect to the extension length, which is closely correlated with the knot size.
These measurements are comparable to the measurement of the variation in $F$ with
knot length carried out in Ref.~\onlinecite{dai2015metastable} for wider channels.

In addition to confinement to channels of constant cross-sectional area, we also examine 
confinement of a knotted polymer to a conical channel.  Conical confinement of polymers has 
been the subject of recent experimental\cite{bell2017asymmetric} and theoretical%
\cite{nikoofard2013directed,nikoofard2015flexible,kumar2018polymer,polson2019polymer}
work, though to our knowledge the effects of such confinement on knots has not yet been
considered.  Here, we examine the variation of the free energy with respect to knot 
position along the channel for a polymer tethered at the narrow end of the cone. For 
convenience, we also choose cone angles that are sufficiently small for Odijk scaling to 
hold throughout.  The variation of the free energy functions with cone angle and knot 
complexity can be understood in the context of a theoretical model that is similar in 
spirit to those developed in Refs.~\onlinecite{nakajima2013localization} and 
\onlinecite{dai2015metastable} to describe knots in other scaling regimes.

%\hspace*{0.2in}
\section{Model}
\label{sec:model}

The simulations examine a semiflexible polymer confined to a long, narrow channel.
We model the polymer as a chain of $N$ hard spheres, each with diameter $\sigma$. 
The pair potential for non-bonded monomers is thus $u_{\rm{nb}}(r)=\infty$ for $r\leq\sigma$ 
and $u_{\rm{nb}}(r)=0$ for $r>\sigma$, where $r$ is the distance between the centers of the 
monomers. Pairs of bonded monomers interact with a potential $u_{\rm{b}}(r)= 0$ if 
$0.9\sigma<r<1.1\sigma$ and $u_{\rm{b}}(r)= \infty$, otherwise. Thus, the length of each
bond fluctuates slightly about its average  value.
The bending rigidity of the polymer is modeled using a bending potential with the form, 
$u_{\rm bend}(\theta)=\kappa(1 - \cos\theta)$. The angle $\theta$ is defined for a consecutive 
triplet of monomers centered at monomer $i$ such that $\cos\theta_{i}=\hat{u}_{i}\cdot\hat{u}_{i+1}$, 
where $\hat{u}_{i}$ is the unit vector pointing from monomer $i-1$ to monomer $i$. The bending 
constant $\kappa$ determines the overall stiffness of the polymer and is related to the 
persistence length $P$ by\cite{micheletti2011polymers} $\exp(-\langle l_{\rm bond} \rangle/P) 
= \coth(\kappa/k_{\rm B}T) - k_{\rm B}T/\kappa$. For our model, the mean bond length is 
$\langle l_{\rm bond} \rangle \approx \sigma$. For sufficiently large $\kappa/k_{\rm B}T\gg 1$
this implies $P/\sigma\approx \kappa/k_{\rm B}T$.

\begin{figure}[!ht]
\begin{center}
\vspace*{0.2in}
\includegraphics[width=0.4\textwidth]{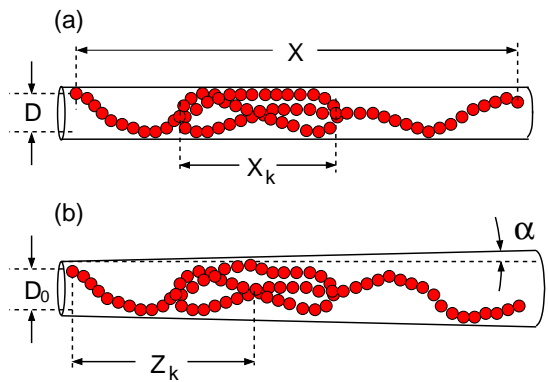}
\end{center}
\caption{Illustration showing the definition of the various parameters described in
the text for a polymer confined to (a) a cylindrical channel and (b) a conical channel.  
The polymer extension, $X$, and knot span, $X_{\rm k}$ shown in panel~(a) are defined
exactly the same for the system in panel~(b).}
\label{fig:model_illust}
\end{figure}

In most simulations, the confining channel is a hard cylindrical tube of uniform  diameter 
$D$.  Each monomer interacts with the wall of the cylindrical tube with a potential
$u_{\rm w}(r) = 0$ for $r<D/2$ and $u_{\rm w}(r) = \infty$ for $r>D/2$, where $r$ is the
distance of the monomer center from the central axis of the cylinder. Thus, $D$ is defined
to be the diameter of the cylindrical volume accessible to the centers of the monomers and
the actual diameter of the cylinder is $D+\sigma$. A second confinement geometry that we 
examine is a hard conical channel with nonuniform diameter $D(z,\alpha)=D_0+2z\tan\alpha$. 
Here, $z$ is the distance along the channel axis and $\alpha$ is the half-angle of the cone. 
In this case, we fix one end monomer to position $z$=0, where the diameter is $D(0,\alpha)$=$D_0$.
The various parameters describing the two model systems are illustrated in Fig.~\ref{fig:model_illust}.

In most cases the confined polymer contains a single knot of topology $3_1$, $4_1$, or $5_1$. 
We also consider the case of an unknotted polymer, for which an S-loop containing two 
hairpin turns may be present for sufficiently short end-to-end polymer extension length.
In order to maintain the knot topology of the polymer we constrain the hairpin turns 
of the knot or S-loop to lie between the two end monomers along the channel. In effect, 
the knot or S-loop is constrained to lie between two virtual walls attached to the end 
monomers that slide along the channel with these monomers. Generally, this feature introduces 
only very weak artifacts in the free energy and other data, as discussed in Sec.~\ref{subsec:cylin}.
	
\section{Methods}
\label{sec:methods}

Monte Carlo simulations are used to calculate the configurational free energy $F$ of the confined
polymer. In the case of a polymer confined to a cylinder, we calculate $F$ as a function
of $X$, the end-to-end extension length of the polymer along the channel axis. In addition,
we examine the $X$-dependence of the knot span, $X_{\rm k}$, which is defined as the distance 
measured along the channel between the tips of the two hairpin turns present in the knot or
S-loop.  For a polymer confined to a conical channel, we measure $F$ as a function 
of $Z_{\rm k}$, the position of the center of the knot along the channel axis measured 
with respect to the end monomer fixed at the narrow end of the channel. The quantities $X_{\rm k}$
and $Z_{\rm k}$ are both illustrated in Fig.~\ref{fig:model_illust}.

To measure the free energy functions, the simulations employed the Metropolis algorithm 
and the self-consistent histogram (SCH) method.\cite{frenkel2002understanding} The SCH 
method can be used to find the free energy function $F(\lambda)$, where $\lambda$ is any 
quantity that is a function of the monomer coordinates. In this study, we choose
$\lambda$=$X$ for cylindrical confinement and $\lambda$=$Z_{\rm k}$ for conical confinement.
To implement the method we carry out many independent simulations, each of which employs a 
unique ``window potential'' of the form:
\begin{equation} 
W_i(\lambda) =
\begin{cases}
 \infty,  \hspace{8mm} \lambda > \lambda_i^{\rm max} \cr
 0 ,      \hspace{9mm} \lambda_i^{\rm min} \leq \lambda \leq \lambda_i^{\rm max} \cr
 \infty,  \hspace{8mm} \lambda < \lambda_i^{\rm min}
\end{cases}
\label{eq:winpot}
\end{equation}
where $\lambda_i^{\rm min}$ and $\lambda_i^{\rm max}$ are the limits that define 
the range of $\lambda$ for the $i$-th window.  Within this range, a
probability distribution $p_i(\lambda)$ is calculated in the simulation. The window 
potential width, $\Delta \lambda \equiv \lambda_i^{\rm max} - \lambda_i^{\rm min}$, 
is chosen to be sufficiently small that the variation in $F$ does not exceed 
2--3 $k_{\rm B}T$.  The windows are chosen to overlap with half of the adjacent 
window, such that $\lambda^{\rm max}_{i} = \lambda^{\rm min}_{i+2}$.  The window width 
was typically $\Delta \lambda = 2\sigma$.  The SCH algorithm was employed to reconstruct 
the unbiased distribution, ${\cal P}(\lambda)$ from the $p_i(\lambda)$ histograms.  
The free energy follows from the relation $F(\lambda) = -k_{\rm B}T\ln {\cal P}(\lambda)$.
A detailed description of the implementation of the SCH algorithm for a polymer system 
comparable to that studied here is presented in Ref.~\onlinecite{polson2013simulation}.

For the case of a knotted polymer confined to a conical channel, calculation
of the free energy energy function $F(Z_{\rm k})$ required a more advanced approach
than a straightforward application of the multiple-histogram method used for $F(X)$
in the case of cylindrical channels. The problem is due to long correlation times
associated with fluctuations in the polymer extension and, correspondingly, in the 
knot span, $X_{\rm k}$. Typically, the correlation times are comparable to the run
time of an entire simulation. The variation of $F$ with $Z_{\rm k}$ was found to
depend significantly on the knot span. Consequently, the histogram associated with
each window in Eq.~(\ref{eq:winpot}) are sensitive to the initial values of $X_{\rm k}$,
which randomly distributed in the initialization routine. This tended to result in
free energy functions of poor quality. To address this problem, we use the 
multiple-histogram method to measure $F(Z_{\rm k})$ for fixed $X$ (which essentially 
also fixes the knot span), and then carry out an appropriate average of these functions 
for a collection of values of $X$.  The details of the method are outlined in 
Appendix~\ref{app:a}.

Polymer configurations were generated by carrying out single-monomer moves using a 
combination of translational displacements and crankshaft rotations. In addition,
standard reptation moves were also employed for the case of cylindrical confinement.
The maximum values of the displacements and rotations are chosen to be small enough 
to not alter the knot topology of the polymer. Trial moves were accepted with a probability 
$p_{\rm acc}$=${\rm min}(1,e^{-\Delta E/k_{\rm B}T})$, where $\Delta E$ is the difference
in the total energy between trial and current states. Note that $\Delta E$=$\infty$ if
any nonbonded monomers overlap, or if the bonding constraints or window potential
constraints of Eq.~(\ref{eq:winpot}) are violated, in which case $p_{\rm acc}=0$ and
the move is rejected. Otherwise, $\Delta E$ is simply the difference in the total
bending energy.  Simulations for polymers confined to a
cylinder employed a polymer of length $N$=400 monomers. The calculations of $F(Z_{\rm k})$
for confinement in a conical channel used shorter chains of $N$=200 monomers because
of the much larger number of simulations required for each free energy function.
Equilibration times were chosen to be sufficiently long to ensure the decay of transients 
in measured quantities that arise from artificial (though convenient) initial configurations.
The system was equilibrated for typically $5\times 10^6$ MC cycles, following which a 
production run of $2\times10^8$ MC cycles was carried out. A MC cycle is defined
as a sequence of $N+1$ trial moves, each of which is either a reptation move or else a 
change in the coordinates of a single randomly selected monomer.  The probability of attempting
a reptation move was chosen to be equal to that of moving any single monomer. For single-monomer 
movement, random displacement and rotation are selected with equal probability.

In the simulations we measure both the span and position of the knot or S-loop along the 
confining channel. This requires identifying the portion of the polymer that is contained
within the knot. One option is to compute the Alexander polynomial of the chain after
closing both ends by a loop based on the minimally interfering closure scheme.\cite{tubiana2011probing}
Although this method is widely used in simulation studies of knotted polymers,
we have chosen not to adopt this approach in the present study. The central problem is 
that the calculations of the variation of $F$ with knot position require that the the knot 
lie within the range of the window defined by the potential of Eq.~(\ref{eq:winpot}). 
Consequently, determining whether a MC move is accepted or rejected requires calculation 
of the knot position each time a move is attempted. The high computational cost of the 
chain-closure method makes this approach infeasible. Fortunately, the fact that we consider 
knotted polymers confined to very narrow channels provides a pragmatic alternative. In this
regime the knot is characterized by two hairpin turns, and the presence of additional
hairpins for polymer contour lengths considered here is extremely improbable.  It is 
straightforward to determine the positions of the monomers at the hairpins with minimal 
computational cost.  We define the span of the knot (or S-loop, in the case of an unknotted
polymer) as the distance along the channel between the these two hairpins and its position 
as the mean position of the hairpins. This approach is similar to that employed by 
M\"{o}bius {\it et al.}, who studied unknotting kinetics for a polymer confined to
a channel under Odijk conditions.\cite{mobius2008spontaneous}

For the results presented below, distances are measured in units of $\sigma$ 
and energies are measured in units of $k_{\rm B}T$. 

\section{Results}
\label{sec:results}

\subsection{Cylindrical channels}
\label{subsec:cylin}

\begin{figure}[!ht]
\begin{center}
\vspace*{0.2in}
\includegraphics[width=0.45\textwidth]{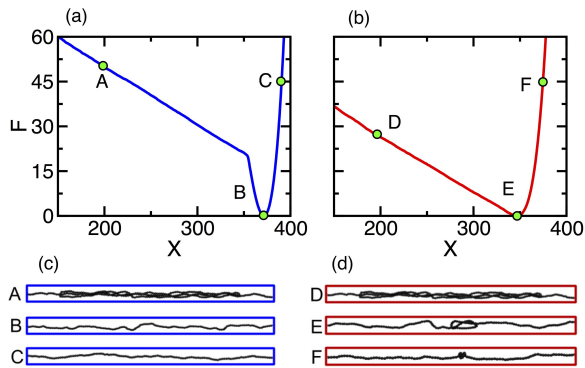}
\end{center}
\caption{(a) Free energy $F$ vs extension length $X$ for a polymer of length $N=400$ 
and a bending constant of $\kappa=15$, confined within a cylindrical channel of diameter 
$D$=$4$. The polymer contains an S-loop in the linear regime.  (b) As in panel (a) except 
for a polymer with a trefoil knot present instead of an S-loop.  (c) Sample conformations 
corresponding to the points labeled in panel (a) for the S-loop. (d) Sample conformations 
corresponding to the points labeled in panel (b) for the trefoil knot.  }
\label{fig:Fillust}
\end{figure}

We first examine the properties of the free energy function $F(X)$ for polymers confined
to a cylindrical channel. Figures~\ref{fig:Fillust}(a) and (b) show representative 
functions for an unknotted polymer and a polymer with a single $3_1$ knot, respectively. 
For $X\lesssim 350$, the unknotted polymer is buckled and contains an S-loop.
(An S-loop is an unknotted structure containing at least two hairpin turns and three elongated
subchains between the hairpins, similar in appearance to the knot structure shown in 
Fig.~\ref{fig:model_illust}(a) except for the key difference in topology.)
In each case, results are shown for a polymer of length $N$=400, bending rigidity of
$\kappa$=15, and confining cylinder diameter of $D$=4. Sample snapshots of the 
polymer at three different extension lengths identified by the three points labeled 
in each graph are shown in panels (c) and (d). The free energy functions share 
common features of (i) the presence of a single minimum at large $X$, (ii) a broad linear
regime at lower $X$, and (iii) a steep rise in the free energy at the highest
extensions. The slopes of the curves in the linear regime are nearly equal for the
two systems.  

The key qualitative difference between the functions is the deep free energy well around 
the minimum present in the case of the unknotted polymer (labeled point B in 
Fig.~\ref{fig:Fillust}(a)). The origin and scaling properties of this free energy 
well have been explained previously.\cite{polson2017free} 
The extension at the free energy minimum corresponds roughly 
to the mean extension length for an elongated semiflexible polymer in the Odijk
regime, where no backfolding is present. Upon decreasing the end-to-end
extension $X$, the polymer buckles and eventually forms two hairpin turns that
constitute the S-loop. The depth of the well is a measure of the free energy 
associated with the formation of the hairpins. Further decreasing $X$ increases
the span of the S-loop along the channel, but leaves the hairpins unaffected.
The linear increase of $F$ with decreasing $X$ arises from the interactions between
the three subchains of the S-loop that lie between the two hairpins. For sufficiently
narrow channels the scaling of the free energy gradient in the linear regime, 
$f\equiv dF/dX$, is expected\cite{odijk2008scaling} to scale with $D$ and the persistence 
length, $P$, according to $f\sim D^{-5/3}P^{-1/3}$, with small deviations in the scaling 
exponents arising from finite-size effects.\cite{polson2017free,polson2018free} At 
higher extensions, where $X>X_{\rm min}$, the rapid increase in $F$ arises from the decrease 
in entropy associated with the suppression of lateral fluctuations in the conformations 
sampled by the polymer.

\begin{figure}[!ht]
\begin{center}
\vspace*{0.2in}
\includegraphics[width=0.45\textwidth]{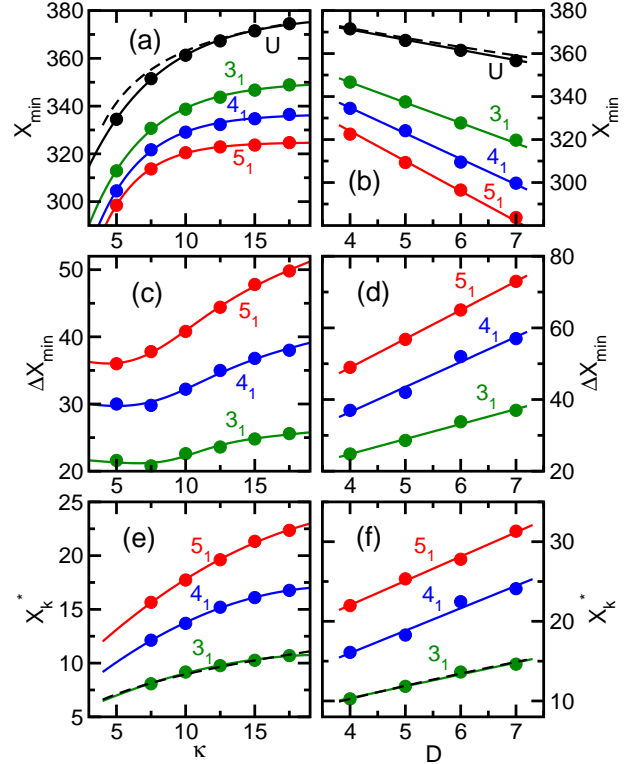}
\end{center}
\caption{(a) Extension length at the free energy minimum $X_{\rm min}$ vs bending rigidity 
$\kappa$ for a polymer of length $N$=400 in a tube of diameter $D$=4. Results are shown for 
unknotted (U) polymers and polymers with a single knot of $3_1$, $4_1$ and $5_1$ topology. 
The solid lines are guides for the eye. The dashed line is the Odijk prediction for the
equilibrium extension length.  (b) As in (a), except $X_{\rm min}$ vs $D$ for polymer bending 
rigidity of $\kappa$=15.  (c) $\Delta X_{\rm min}$ vs bending rigidity $\kappa$ for $D$=4, 
where $\Delta X_{\rm min}$ is defined in the text. (d) As in (c), except $\Delta X_{\rm min}$ 
vs $D$ for polymer bending rigidity of $\kappa$=15.  (e) Knot span $X_{\rm k}^*$ at 
$X=X_{\rm min}$ vs $\kappa$ for $D$=4. (f) Knot span $X_{\rm k}^*$ at $X=X_{\rm min}$ 
vs $D$ for $\kappa$=15. The dashed black lines in panels (e) and (f) are the functions
$1.65 D^{2/3}P^{1/3}$ for fixed $D$=4 and $P$=15, respectively.  }
\label{fig:Xmin.delXmin}
\end{figure}

Unlike the case for an unknotted polymer, the hairpins present in a knotted polymer are not 
eliminated when the extension length increases and the polymer unbuckles. Thus, there is no 
corresponding release of the hairpin free energy (which is mainly the hairpin bending energy 
for a polymer in the Odijk regime) when an S-loop is removed. Consequently, the deep free energy 
well associated with the hairpin formation is not present.  Obviously, the value of $X_{\rm min}$ 
is closely connected to both the most probable knot contour length and knot span length. The 
greater each of these lengths are, the lower the corresponding value of $X_{\rm min}$. As will 
be examined in detail below, the values of these quantities are each strongly affected by the 
polymer bending rigidity, the channel diameter, and the topology of the knot. Note that the 
scaling properties of the free energy for knotted polymers confined to 
channels have been elucidated in two previous studies;\cite{nakajima2013localization,%
dai2015metastable} however, neither approach is directly applicable to interpreting the 
present results.  Ref.~\onlinecite{nakajima2013localization} considered fully-flexible polymers 
in the de~Gennes regime, while Ref.~\onlinecite{dai2015metastable} examined knots in 
semiflexible polymers, they used channels of width $D\gtrsim P$, which is wider than those 
considered here. Those studies considered only trefoil knots, and in each of them a metastable 
knot was observed whose most probable size was dependent on the channel dimension. The underlying 
factors governing the scaling behavior of the typical size of a trefoil knot in the Odijk regime 
will be examined later in this section of the article in the analysis of the data of 
Fig.~\ref{fig:Xmin.delXmin}(e) and (f).

\begin{figure}[!ht]
\begin{center}
\vspace*{0.2in}
\includegraphics[width=0.45\textwidth]{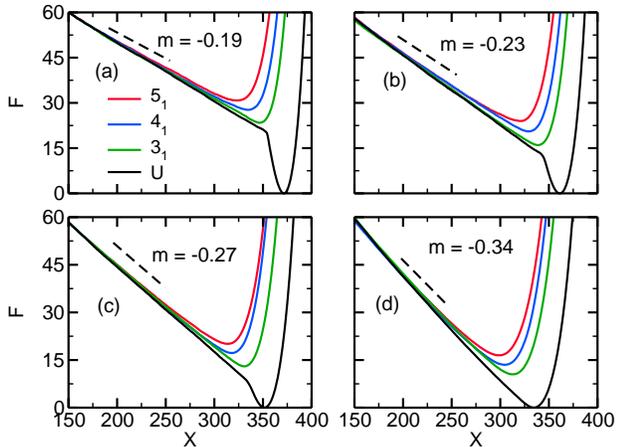}
\end{center}
\caption{Free energy $F$ vs extension length $X$ for both knotted polymers ($3_1$, $4_1$, 
and $5_1$) and unknotted polymers that may contain an S-loop (U). The polymers are of length 
$N$=400 and are confined to a cylindrical channel with a diameter of $D$=4. Results are shown
for (a) $\kappa$=$15$, (b) $\kappa$=$10$, (c) $\kappa$=7.5, and (d) $\kappa$=5. In each panel,
the slope $m$ obtained obtained for a fit to the linear portion of $F$ for the S-loop is
labeled.}
\label{fig:F.kappa.D=4}
\end{figure}

Figure~\ref{fig:F.kappa.D=4} shows free energy functions for polymers of length $N$=400
in narrow cylindrical channels with $D$=4. Results are shown for bending rigidities 
in the range $\kappa=5-15$, and each panel shows functions for a given value of $\kappa$
for unknotted polymers, as well as those with knots with topologies of $3_1$, $4_1$ and $5_1$. 
In the case of an unknotted polymer, the depth of the free energy well decreases as the polymer 
becomes more flexible. This is mainly due to the reduction in the hairpin bending energy 
that is released as the extension $X$ increases and the polymer unbuckles. Indeed, at $\kappa$=5 no
free energy well is present, as expected for the regime $D\gtrsim P$ where the concept 
of a hairpin turn is no longer meaningful. The slope of the curves in the linear regime
gradually increases as the polymer rigidity lessens. This is qualitatively consistent with 
the theoretical prediction that the slope scales as $P^{-1/3}$ in the Odijk regime.%
\cite{odijk2008scaling,polson2017free,polson2018free} Another notable trend is the overlap in
the curves for different topologies at each $\kappa$ in the linear regime. This overlap
is not perfect, but close enough to suggest that polymer knot topology does not strongly affect 
the overall conformational behavior of the knot/S-loop when that structure has a sufficiently
large span along the channel. As a clarifying example, the conformational behavior illustrated
in the snapshots of state A in Fig.~\ref{fig:Fillust}(c) for an S-loop and state D in 
Fig.~\ref{fig:Fillust}(d) for a $3_1$ knot are, by this measure, very similar. 
As $X$ increases, the free energy curve for each knotted polymer eventually peels away
from the S-loop curve. Generally, the value of $X_{\rm min}$ for each knot topology 
decreases with increasing complexity of the knot. Thus, the most probable contour length 
and knot span increases with knot complexity.

\begin{figure}[!ht]
\begin{center}
\vspace*{0.2in}
\includegraphics[width=0.45\textwidth]{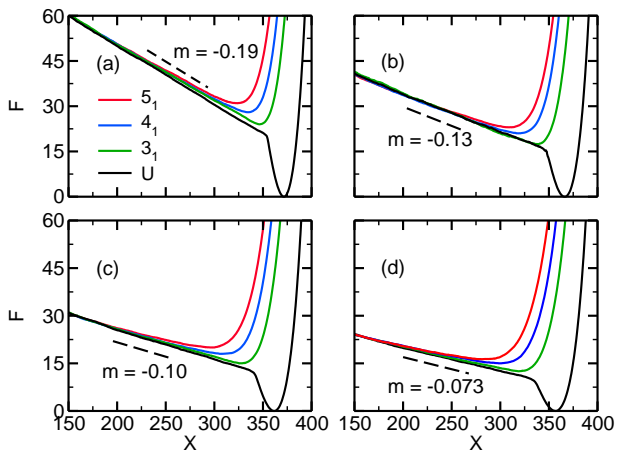}
\end{center}
\caption{Free energy functions for knotted polymers ($3_1$, $4_1$, and $5_1$) and unknotted 
polymers that may contain an S-loop (U). Each polymer has a length of $N$=400 and a bending 
rigidity of $\kappa$=15. Results are shown for (a)$D$=4, (b) $D$=5, (c) $D$=6, and (d) $D$=7. }
\label{fig:F.D.kappa=15}
\end{figure}

Figure~\ref{fig:F.D.kappa=15} shows free energy functions for a polymer of length $N$=400
with a fixed bending rigidity of $\kappa$=15. Results are shown for cylinder diameters
ranging from $D=4-7$ for each of the topologies considered in Fig.~\ref{fig:F.kappa.D=4}.
Again, we note the approximate overlap in the linear regime between the curves for knots with 
different topologies. As before, this suggests that knots and S-loops have similar conformational
behavior in the case where these structures are sufficiently large. The value of the slope 
decreases with increasing channel width. This is qualitatively consistent with the 
expectation that the slope scales as $D^{-5/3}$ in the Odijk regime.\cite{polson2017free,%
polson2018free} As this slope lessens with increasing $D$, there is a widening in the 
distribution of extension lengths resulting from an increase in the probability of 
shorter extension lengths. This corresponds to a widening of the knot size distribution
through an increase in the probability of larger knots.
This is qualitatively consistent with the trend observed in simulations of Jain and Dorfman
for knotted polymers in square channels for confinement near the onset of Odijk scaling
(i.e. $D\approx P$).\cite{jain2017simulations} As in Fig.~\ref{fig:F.kappa.D=4}, 
$X_{\rm min}$ decreases as the complexity of the knot topology increases. Thus, the most 
probable contour length and span of the knot increases with knot complexity.

As noted in Sec.~\ref{sec:model}, the knot or S-loop is artificially constrained to lie 
completely between the two ends of the polymer. This feature was incorporated into the model 
to prevent the polymer knot from untying or changing to a different knot type. This artificial 
confinement is expected to reduce the entropy of the system and thus increase the free energy. 
Here, we estimate the effect on the free energy functions by modeling the knot as a 
particle undergoing a 1-D random walk along the channel. For a channel of constant 
cross-sectional area, the energy is independent of the knot position. Neglecting
fluctuations in the span of the knot, the range of positions accessible to its center is
$X-X_{\rm k}$. Thus, the entropy is $S_{\rm c}/k_{\rm B}=\ln(X-X_{\rm k}) + {\rm const.}$,
and the free energy is $F_{\rm c}/k_{\rm B}T = -S_{\rm c}/k_{\rm B} = -\ln(X-X_{\rm k})+{\rm const.}$
Note that the knot span $X_{\rm k}$ depends on $X$, $D$ and $\kappa$, but is insensitive to
knot topology for extensions in the linear regime of the free energy (i.e. $X$ is sufficiently
less than $X_{\rm min}$).

\begin{figure}[!ht]
\begin{center}
\vspace*{0.2in}
\includegraphics[width=0.40\textwidth]{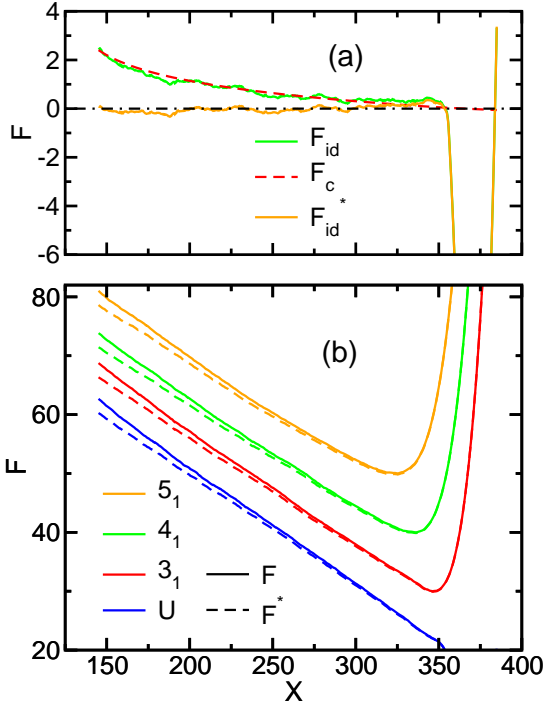}
\end{center}
\caption{(a) Illustration of the effect of artificial longitudinal confinement of the knot/S-loop on the
free energy. $F_{\rm id}$ is the free energy function of an ideal polymer (i.e. overlap of monomers
is permitted) for $N$=400, $D$=4, and $\kappa$=15.  The longitudinal confinement free energy is
defined: $F_{\rm c}\equiv -\ln(X-X_{\rm k})+{\rm constant}$, where $X_{\rm k}=X_{\rm k}(X)$ is
the knot span. The shifted free energy is $F^*_{\rm id}\equiv F_{\rm id}-F_{\rm c}$. (b) Free 
energy for real polymers (i.e. no overlap between monomers). Note that $F^*\equiv F-F_{\rm c}$. 
The polymer has either a knot or an S-loop. As in (a), $N$=400, $D$=4, and $\kappa$=15.  }
\label{fig:F.correct}
\end{figure}

To test this approximation, we calculate $F(X)$ for an ``ideal'' polymer,
by which we mean that monomer-monomer overlap is permitted. Note that the topology of the polymer
will not be preserved in the simulation, but this is not expected to matter for $F_{\rm c}$.
Figure~\ref{fig:F.correct}(a) shows the free energy function $F_{\rm id}$ for an ideal polymer
with $N$=400, $D$=4 and $\kappa$=15. As expected, $F_{\rm id}$ does not display the steep
increase with decreasing $X$ seen in Fig.~\ref{fig:F.kappa.D=4}(a) for a ``real''
polymer system (i.e. where no monomer-monomer overlap is permitted) with otherwise the
same conditions.  However, there is a residual small increase in $F$ with decreasing $X$ 
resulting from the artificial confinement described above.
Overlaid on this curve is the estimate of $F_{\rm c}$. We see
excellent agreement between the two results in the regime where the two hairpins are
present (i.e. $X<350$).  The corrected free energy, $F_{\rm id}^*\equiv F_{\rm id}-F_{\rm c}$
is now independent of polymer extension, demonstrating the validity of the approximation.
Figure~\ref{fig:F.correct}(b) shows free energy functions $F(X)$ and corrected free energy
functions, $F^*(X)$, where the latter are calculated as in Fig.~\ref{fig:F.correct}(a).
The correction leads to a small but significant change in the curves. Specifically,
it slightly decreases the free energy gradient in the linear regime and extends
the range of $X$ over which the curves remain linear.

The variation of $X_{\rm min}$ for the free energies of Figs.~\ref{fig:F.kappa.D=4} and 
\ref{fig:F.D.kappa=15} with $\kappa$ and $D$ for both knotted and unknotted polymers 
is shown in Fig.~\ref{fig:Xmin.delXmin}(a) and (b), respectively. As expected, the Odijk 
prediction for the mean extension length, which is overlaid on the data, agrees well with 
the results for the unknotted polymer, particularly at large $\kappa$ and low $D$.
To quantify the shift in $X_{\rm min}$ of the knotted polymers relative to that of the 
unknotted polymer, we define $\Delta X_{\rm min}\equiv X_{\rm min}(n_1)-X_{\rm min}(U)$, 
where $X_{\rm min}(n_1)$ is the extension length at the free energy minimum for a polymer 
with a knot of topology $n_1$ for $n$=3, 4, and 5, and $X_{\rm min}(U)$ is the corresponding 
extension length of an unknotted polymer. $\Delta X_{\rm min}$ a measure of the reduction in 
the extension length of the polymer caused by the presence of the knot and is roughly 
proportional to the contour length of the knot.  Figures~\ref{fig:Xmin.delXmin}(c) 
and (d) shows the variation of $\Delta X_{\rm min}$ with $\kappa$ and $D$, respectively. 
A clearer measure of knot size is $X_{\rm k}^*$, span of the knot along the channel evaluated 
at the free energy minimum, $X=X_{\rm min}$. The knot span is defined as the distance measured 
along the channel between the tips of the two hairpins in the knot. Figures~\ref{fig:Xmin.delXmin}(e) 
and (f) show the variation of $X_{\rm k}^*$ with $\kappa$ and $D$, respectively. 
As expected, the general trends for $\Delta X_{\rm min}$ are the same as for $X_{\rm k}^*$, 
since both are measures of knot size.  For each measure of knot size, three main trends are 
apparent. First, $\Delta X_{\rm min}$ and $X_{\rm k}^*$ increase with both the polymer rigidity 
and the channel diameter.  Second, the rate of increase of each with $\kappa$ and $D$ appears 
to increase with increasing knot complexity. Finally, for any given value $\kappa$ and $D$, 
$\Delta X_{\rm min}$ and $X_{\rm k}^*$ increases with knot complexity. An increase in knot 
size with knot complexity for elongated polymers was also observed in the case of an unconfined 
knotted polymer under tension,\cite{caraglio2015stretching} as well as for polymers confined
to channels somewhat wider than those examined here (i.e. $D\approx P$).\cite{jain2017simulations}
However, the increase in knot size with increasing $D$ differs from the behavior 
observed in Ref.~\onlinecite{jain2017simulations}, where it remained relatively unchanged.

What factors determine the most probable knot span? The rapid rise in 
$F$ at high $X$ corresponds mainly to the loss in entropy associated with the suppression 
lateral conformational fluctuations. In addition, increasing $X$ in the linear regime 
($X<X_{\rm min}$) leads to a reduction in knot size. For a sufficiently large knot
this reduces excluded volume interactions between deflection segments in the knot 
and causes the decrease in $F$. These two contributions to $F$ alone guarantee the 
presence of a minimum. Now consider further the intra-knot excluded volume interactions. 
The prediction\cite{odijk2008scaling} and subsequent verification by computer simulation%
\cite{polson2017free} that the free energy
gradient of an S-loop approximately scales as $f\equiv dF/dX \sim D^{-5/3}P^{-1/3}$ is derived by
modeling the polymer as a collection of equivalent hard cylinders. The cylinders have
a length given by the Odijk deflection length $l_d\sim D^{2/3}P^{1/3}$ and inter-cylinder 
interactions are estimated using the second-virial approximation. In this picture, increasing 
$X$ corresponds to shortening the knot and removing these virtual hard cylinders out of the 
knot into a region where no such interactions are present. Thus, $F$ decreases. Eventually,
however, when the knot span is of the order of $l_{\rm d}$, this picture breaks down.
The strands between the hairpins are (obviously) connected to the hairpins. These
constraints are expected to severely constrain the orientational freedom of the (effectively
rigid) strands, in a manner that the orientational entropy sharply drops with shortening 
knot span. Thus, it is expected that when $X_{\rm k}\approx l_{\rm d}$ a contribution to the 
free energy emerges that steeply rises with increasing $X$, leading to a minimum in the 
free energy.  

The simple argument above suggests that $X_{\rm k}^*\approx l_{\rm d}$.
The dashed curves in panels (e) and (f) of Fig.~\ref{fig:Xmin.delXmin} are plots of
$1.65D^{2/3}P^{1/3}$ for for fixed $D$=4 and $P$=15, respectively. These dashed curves 
overlap with the $X_{\rm k}^*$ perfectly, suggesting this argument is valid for trefoil
knots. On the other hand, attempts to fit the data for $4_1$ and $5_1$ knots using this 
scaling were not successful. The increased entanglement for elongated knots of greater
topological complexity likely introduces additional constraints for those knots that
further reduce the orientational freedom of the knot strands. Clearly, this effect kicks in
at larger knot span that is not simply a multiple of $l_{\rm d}$. Further elucidation
of such effects in a future study would be worthwhile.

\begin{figure}[!ht]
\begin{center}
\vspace*{0.2in}
\includegraphics[width=0.45\textwidth]{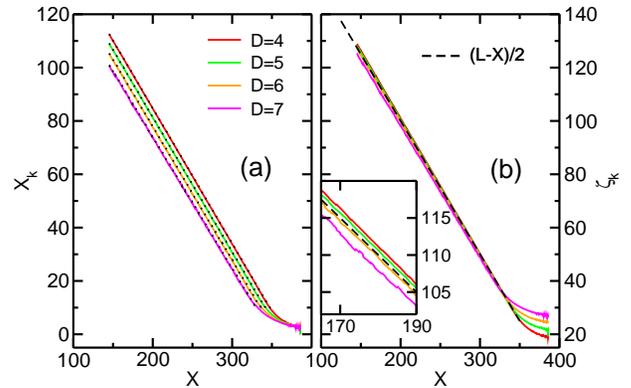}
\end{center}
\caption{(a) Variation of the extension length of a $3_1$ knot with respect to the
extension length of a polymer confined to a channel. Results are shown for a
polymer of bending rigidity $\kappa$=15 in a channel of various channel diameters.
The black dotted lines overlaid on the solid curves are corresponding results for
the S-loop extension length for an unknotted polymer.  (b) Variation of $\zeta_{\rm k}$ 
with $X$, where $\zeta_{\rm k}$ is defined by Eq.~(\ref{eq:Xkdag}) and calculated using 
the data for the $3_1$ knot in panel (a). The dashed line shows the theoretical prediction 
that $\zeta_{\rm k}=(L-X)/2$. The inset shows a close-up of the data.}
\label{fig:kspan}
\end{figure}

Let us now consider the relationship between knot span and polymer extension length.
Figure \ref{fig:kspan}(a) shows the variation in the knot span along the channel with the
extension length of a polymer with a $3_1$ knot. Results are shown for a polymer of
bending rigidity $\kappa$=15 and for various channel diameters. The black dotted curves 
overlaid on the data are corresponding results for the span of an S-loop for an
unknotted polymer. Several trends are apparent. As expected, the knot span decreases
with increasing $X$. For the range of $X$ corresponding to the linear regime in
the free energy functions of Fig.~\ref{fig:F.kappa.D=4}, $X_{\rm k}$ decreases linearly
with $X$. As $X$ approaches and then passes the extension at minimum free energy, 
$X_{\rm min}$, the rate of decrease of knot span with $X$ decreases. For any value
of $X$, the knot span decreases slightly with increasing channel diameter. However,
for each $D$, the curves are essentially parallel with a slope $dX_{\rm k}/dX\approx -0.5$. 
In the case of an unknotted polymer for extensions where an S-loop is present, the
S-loop span is virtually identical to the $3_1$ knot span at any $X$.

To understand the origin of these trends, we employ the scaling properties of a 
channel-confined polymer in the Odijk regime. Note that the required condition 
$D\ll P$ is only marginally satisfied in this case ($P/D = 2.14-3.75$), and thus 
some quantitative discrepancy between the predicted and observed behavior is to be
expected. Since the results for knotted ($3_1$) and unknotted (S-loop) polymers
are identical over the regime of interest ($X<X_{\rm min}$), we ignore the effects 
of topology. Let us first consider a polymer with no backfolding. Recall that the 
mean extension length of a polymer in this regime is given by 
$\bar{L}_{\parallel}=L(1-\alpha_{\parallel} D^{2/3}P^{-2/3})$, where $L$ is the 
contour length of the polymer and where the prefactor is 
$\alpha_{\parallel}=0.1701\pm 0.0001$.\cite{dai2016polymer} 
Now, consider a polymer with two hairpin folds, which may result from an S-loop or a knot. 
We first define an effective contour length as $\ell\equiv L-\pi D$ to exclude the contour 
in the two hairpins.  Here, we assume the hairpin diameter is $D$, which is likely only a 
slight overestimate in the narrow-channel limit.\cite{odijk2006dna,chen2017conformational}
The mean span of the S-loop/knot, ${X}_{\rm k}$, is the mean distance between the two 
hairpins. We next define the effective extension of the polymer as the sum of the
extensions along the channel of all elongated pieces of the polymer, which excludes
the hairpins.  As explained in the Supplemental Material, the effective extension 
of the polymer is given by $\ell_\parallel = 2{X}_{\rm k} + X - 2D$.
Replacing $L\rightarrow \ell$ and $L_\parallel \rightarrow \ell_\parallel$ in
the relation for $L_\parallel$ above, it follows:
\begin{eqnarray}
{X}_{\rm k} & = & -{\textstyle\frac{1}{2}} X + {\textstyle\frac{1}{2}} L 
            - ({\textstyle \frac{\pi}{2}}-1)D
            - {\textstyle\frac{1}{2}} \alpha_\parallel L (D/P)^{2/3} \nonumber \\
            & & + {\textstyle\frac{1}{2}} \alpha_\parallel \pi D (D/P)^{2/3}.
\label{eq:Xk}
\end{eqnarray}
The first term accounts for the observed slope of $dX_{\rm k}/dX\approx -0.5$, while the third
and fourth terms account for the observed decrease in $X_{\rm k}$ with increasing $D$.  
In our simulations, $\pi D\ll L$, and so the 5th term is negligible relative to
the 4th term and can be omitted.  In the calculation above, we have neglected the effects of
fluctuations in the extension length and interactions between the elongated segments
in the knot/S-loop. In the Supplemental Material, we show that these effects are 
negligible.  Defining the shifted knot extension, $\zeta_{\rm k}$, as
\begin{eqnarray}
\zeta_{\rm k} & \equiv & {X}_{\rm k} + ({\textstyle \frac{\pi}{2}}-1)D
            + {\textstyle\frac{1}{2}} \alpha_\parallel L D^{2/3} P^{-2/3},
\label{eq:Xkdag}
\end{eqnarray}
it follows from Eq.~(\ref{eq:Xk}) (omitting the negligible 5th term) that $\zeta_{\rm k}=(L-X)/2$ 
for all values of $D$, $P$, independent of the  polymer extension $X$. Figure~\ref{fig:kspan}(b) 
shows that such a shift 
does lead to near collapse of the data to the predicted curve for $N$=400 in the range of $X$ 
corresponding to the linear regime of the free energy. The data collapse is slightly worse 
for the largest channel diameter of $D$=7, where the Odijk regime conditions are least well 
satisfied.  Overall, the data collapse to a universal curve is reasonably good, given the 
approximations employed in this theoretical model. 

\subsection{Conical channels}
\label{subsec:conical}

We now consider the behavior of a knotted polymer in a conical channel. Rather than 
measuring the free energy with respect to polymer extension we use instead the knot 
position, $Z_{\rm k}$. The central goal here is to characterize the effects of the 
varying channel cross-sectional area at the location of the knot as it samples
different locations along the channel. As noted in Section~\ref{sec:methods}
a problem with measuring $F(Z_{\rm k})$ is the very long correlation time
associated with the fluctuations in the polymer extension length. Consequently,
we choose instead the approach described in Appendix~\ref{app:a}. Essentially, this involves
calculation of $F(Z_{\rm k}|X)$, the variation in the free energy with knot position
for fixed polymer extension $X$, and carrying out a suitable average of these functions.

\begin{figure}[!ht]
\begin{center}
\vspace*{0.2in}
\includegraphics[width=0.45\textwidth]{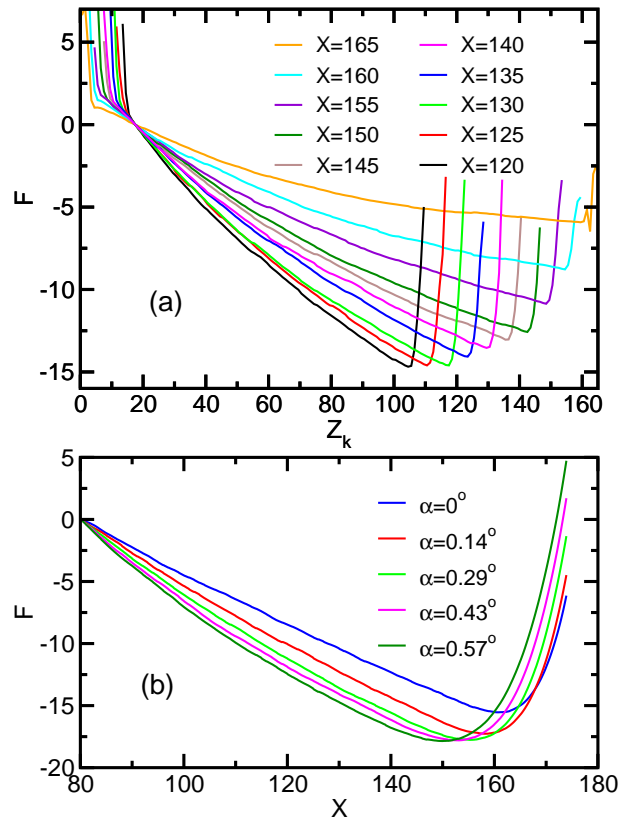}
\end{center}
\caption{(a) Free energy $F$ vs $Z_{\rm k}$ calculated for a polymer under conical
confinement at fixed polymer extension length. The polymer has a length $N=200$, a
bending constant of $\kappa = 15$, and a localized $3_1$ knot along its contour.
The cap of the narrow end of the cone has diameter $D_0=4$, and the cone angle
is $\alpha=0.57^\circ$.  Results for various cone angles are shown. For visual
clarity, the curves are each shifted such that $F$=0 at $Z_{\rm k}$=17.5. 
(b) Free energy $F$ vs extension length for a polymer with $N=200$,
$\kappa = 15$, and a $3_1$ knot confined to a cone with a cap diameter
of $D_0=4$. Results for several cone angles are shown.  }
\label{fig:posext_ang01}
\end{figure}

Figure~\ref{fig:posext_ang01}(a) shows the variation of $F$ with knot position
for a range of polymer extension lengths. Results are shown for a polymer of
length $N$=200 and bending rigidity $\kappa=15$ with a $3_1$ knot confined
to a conical channel with $D_0$=4, and cone angle $\alpha$=0.57$^\circ$.
The steep rise in $F$ at low and high extremes of $Z_{\rm k}$ is an expected 
artifact arising from the constraint that the entire span of the knot lie between 
the two ends of the polymer. (Recall that this constraint is imposed to preserve 
the knot topology and prevent the knot from untying.) This rapid increase
arises when an edge of the knot makes contact with the ``virtual wall'' attached
to an end monomer. This occurs when $|Z_{\rm k}-Z_{\rm end}|\approx X_{\rm k}/2$,
where $Z_{\rm end}$ is the position of the end monomer nearest to the knot center.
As $X$ increases, the knot span decreases and the knot can occupy a wider range of 
positions along the channel before it makes contacts with the virtual wall. 
Thus, as $X$ increases we observe an increase in the distance along the channel 
between these steep increases in $F$. 

In the region where the knot is not close to the end monomers, $F$ decreases 
monotonically as the knot moves in the direction of increasing channel diameter, i.e. 
increasing $Z_{\rm k}$.  In addition, rate of change in $F$ with $Z_{\rm k}$ increases 
monotonically as the extension length increases. The origin of these trends is
straightforward. As the knot moves to a wider part of the channel, the bending
energy associated with the hairpin turns decreases, contributing to a decrease
in $F$. Another contribution to this trend is the free energy associated with the 
overlap of the three strands of the polymer inside the knot between the hairpins, 
which is also expected to decrease as the diameter of the channel decreases. 
This second contribution to the free energy is proportional to the knot extension
length, which decreases as $X$ increases. Consequently, there is a weaker
contribution to variation of $F$ with $Z_{\rm k}$, and thus the rate $dF/dZ_{\rm k}$ 
decreases with increasing extension length.

The method described in Appendix~\ref{app:a} requires calculation of the $F$ with the 
polymer extension length $X$ for knotted polymers confined to a conical channel. 
Results are shown in Fig.~\ref{fig:posext_ang01}(b) for the case of a $3_1$ knot and for 
several cone angles.  For visual clarity, the curves are shifted so that $F$=0 at $X$=80. 
The curves are qualitatively similar to those in Figs.~\ref{fig:F.kappa.D=4} and 
\ref{fig:F.D.kappa=15}. Increasing the cone angle has the effect of increasing
the curvature of the functions for $X<X_{\rm min}$ and decreasing the value of $X_{\rm min}$.
The latter trend results from the fact that the extension length decreases with
increasing $D$ and larger angles correspond to more of the polymer confined to 
wider parts of the channel.

\begin{figure}[!ht]
\begin{center}
\vspace*{0.2in}
\includegraphics[width=0.45\textwidth]{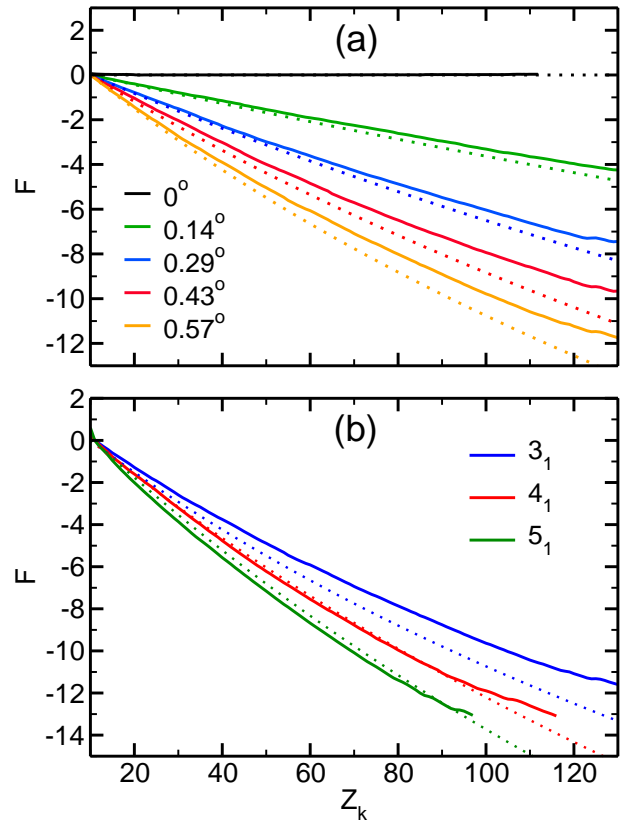}
\end{center}
\caption{(a) Free energy $F$ vs $Z_{\rm k}$ calculated for a polymer under conical 
confinement. The polymer has a length $N=200$, a bending constant of $\kappa = 15$, 
and a localized $3_1$ knot along its contour. The cap of the narrow end of the cone 
has diameter $D_0=4$.  Results for various cone angles are shown. The solid curves are 
the simulation data and the dashed curves are the predictions using the theory developed 
in Appendix~\ref{app:c}. (b) Free energy functions for $N$=200, $\kappa$=15, $D_0$=4, and 
cone angle of $\alpha$=$0.57^\circ$. Results are shown for three different knot topologies.}
\label{fig:theory200}
\end{figure}

Using the method described in Appendix~\ref{app:a} and results such as those shown in 
Fig.~\ref{fig:posext_ang01}, we calculate the variation of the free energy with $Z_{\rm k}$.
Figure~\ref{fig:theory200}(a) shows $F(Z_{\rm k})$ for several cone angles. Results are 
shown for a polymer of length $N$=200 and bending rigidity $\kappa$=15 in a cone with an 
end fixed at a position where the channel diameter is $D_0$=4. We consider cones that deviate 
only slightly from cylindrical channels, with a cone half-angle ranging from $\alpha$=0$^\circ$
(i.e. a cylindrical channel) to $\alpha$=0.57$^\circ$. This range of $\alpha$ is chosen
to ensure that the condition for the Odijk regime is satisfied at least marginally for all 
positions along the channel occupied by the polymer, i.e. $D(z)<P$. The free energy functions 
are shown in the range $10\leq Z_{\rm k}\leq 130$. Inside this range the free energy is 
unaffected by the artificial constraint that the entire span of the knot lie between 
the two end monomers. At either extreme outside this range the knot compresses against
the virtual walls connected to these end monomers and the free energy abruptly rises.
For visual clarity, the free energy curves are shifted so that $F$=0 at $Z_{\rm k}$=10.
Unsurprisingly, the free energy is independent of knot position for cylindrical channels
with constant cross-sectional area. However, for $\alpha>0$ the free energy decreases
monotonically with increasing $Z_{\rm k}$, i.e., as the knot moves to a channel location
with a larger channel diameter. In addition, at any given $Z_{\rm k}$ the decrease
in the free energy relative to the $Z_{\rm k}=10$ reference point is larger for
larger $\alpha$. Thus, the dependence of $F$ on $Z_{\rm k}$ and $\alpha$ indicates
that the knot position probability increases with increasing channel width at the
knot location.
Figure~\ref{fig:theory200}(b) shows free energy functions for three different knots,
each for a polymer with $\kappa$=15 and a cone with $D$=4 and $\alpha$=$0.57^\circ$.
The curves are all qualitatively similar. The key trend is the more rapid decrease
in $F$ with $Z_{\rm k}$ for knots of increasing complexity. 

The trends in Fig.~\ref{fig:theory200} can be better understood using a theoretical description
that incorporates insights gained from recent theoretical studies of folded polymers under
confinement in channels. The theory is developed and described in detail in Appendices~\ref{app:b}
and \ref{app:c}. Here, the channel-confined knotted polymer is modeled as a single linear
polymer of extension $X$ that overlaps with a ring polymer of extension $X_{\rm k}$, 
as illustrated in Fig.~\ref{fig:loopmodel} of Appendix~\ref{app:c}. The ring polymer is
a simple representation of the knot. The lengths of the linear and ring polymer are designed
to vary in a manner such that the total length of the two is held fixed. In this context, 
the free energy has four principal
contributions: (1) the free energy associated with the hairpin turns; (2) the
overlap free energy of the three subchains in the knot that lie between the two hairpins;
(3) the confinement free energy of the extended sections of the ring polymer outside
the hairpins; and (4) the confinement free energy of the linear polymer.
Figure~\ref{fig:Fth_all} in Appendix~\ref{app:c} shows each of the four contributions
to $F(Z_{\rm k})$ for the case of $\kappa=15$, $D_0=4$ and $\alpha=0.57^\circ$.
We note that the dominant contribution is the free energy of the hairpins, which accounts
for about 70\% of the variation of $F$ with knot position. This is mainly a result of the 
large amount of bending energy stored in the hairpins that is released as the channel widens. 
 
Curves for the theoretical predictions of $F(Z_{\rm k})$ are overlaid on simulation results 
in Figs.~\ref{fig:theory200}(a) and (b). The quantitative accuracy of the predictions is 
surprisingly good, given the crudeness of the approximations employed in the theory.
The key qualitative trends are both reproduced by the theory: (1) $F$ decreases more rapidly 
with $Z_{\rm k}$ as the cone angle increases and (2) $F$ decreases more rapidly as the knot 
complexity increases. The first feature is mainly due to the release of the hairpin free energy 
(mainly bending energy) as the knot moves in the direction of increasing channel width. 
The second feature appears to be associated with the increase in knot span with
knot complexity, as well as the greater rate of change of knot size with channel width 
for increasing knot complexity, as observed in Fig.~\ref{fig:Xmin.delXmin}(d) and (f).

\section{Conclusions}
\label{sec:conclusions}

In this study we have used MC simulations to investigate the properties of the conformational
free energy of a knotted semiflexible polymer confined to cylindrical and conical channels.
The channels are sufficiently narrow for the conditions for Odijk scaling ($D<P$) to be 
marginally satisfied. Most other comparable simulation studies of knotted polymers
have considered systems with wider channels corresponding either to extended de~Gennes scaling 
regime or else near the onset of Odijk scaling ($D\approx P$). Those cases are more
relevant to recent experiments of knotted DNA.\cite{metzler2006diffusion,reifenberger2015topological,%
amin2018nanofluidic,ma2020diffusion} Our choice to focus on the Odijk regime is motivated by
an expectation that future experiments for Odijk-regime systems will eventually be carried out, 
as well a basic interest in the fundamental physics of the behavior of knots in a regime that
has been otherwise so thoroughly examined for polymers in the absence of self-entanglement.
This study builds on our recent work of folded semiflexible polymers under confinement%
\cite{polson2017free,polson2018free} and employs similar methodology.

For cylindrical channels, we measured the variation of $F$ with the extension length $X$ 
for polymers with knots of various types, as well as for unknotted polymers. Since the
value of $X$ determines the span of the knot, the calculations in effect measure the
variation of $F$ with knot size. As in other scaling regimes for both flexible\cite{nakajima2013localization}
and semi-flexible\cite{dai2015metastable} chains, we observe a metastable knot, corresponding
to a minimum in $F(X)$. The most probable knot size $X_{\rm k}^*$ increases with persistence length
$P$, channel width $D$, and knot complexity. For trefoil knots, $X_{\rm k}^*$ scales approximately 
with the Odijk deflection length, though the behavior for more complex knots is less straightforward.
For knots in the size regime where $X_{\rm k} > X_{\rm k}^*$ (i.e. knots larger than the most
probable size) the scaling of $F$ with respect to $X$, $P$ and $D$ is comparable to that
for unknotted polymers containing an S-loop. Specifically, in this regime the scaling of
free energy gradient is in approximate agreement with the prediction of $f\equiv dF/dX \sim D^{-5/3}P^{-1/3}$
previously derived\cite{odijk2008scaling} and confirmed\cite{polson2017free} for the case
of an S-loop. In addition, knot span dependence on $X$ and its scaling with $D$ and $P$ is
identical to that of an S-loop. We conclude that the overall conformational behavior of 
knots is very similar to that of an S-loop, at least in the regime where 
$X_{\rm k} > X_{\rm k}^*$.

In addition to cylindrical channels, we also examined the behavior of knots in conical
channels. In this case, we measured the variation of $F$ with respect to knot position
along the channel, $Z_{\rm k}$. Generally, we find that $F$ decreases as $Z_{\rm k}$
increases, i.e., as the knot moves to the wider part of the channel. The main driving
force is the reduction in the hairpin free energy (mainly the hairpin bending energy) with
increasing channel diameter, which is unsurprising given the narrowness of the channels
in the Odijk regime. Generally, we find that the rate of decrease of $F$ with $Z_{\rm k}$ 
increases with increasing cone angle and with knot complexity. A simple theoretical model
that describes the knotted polymer as a linear polymer overlapping with a ring polymer
is able to account for these trends.

One outstanding matter concerns the general criteria that determine the metastable knot
size $X_{\rm k}^*$ for knots of arbitrary complexity and how $X_{\rm k}^*$ scales with respect 
to channel width and persistence length. A future goal in subsequent work will be to develop
a theoretical model for the free energy in the spirit of that developed in 
Ref.~\onlinecite{dai2015metastable} applicable to the Odijk regime and for arbitrary knot type.
The observation that scaling $X_{\rm k}$ matches that of the Odijk deflection length in the 
case of trefoil knots is a useful starting point. In addition, it will be useful to measure
directly the variation of $F$ with $P$ and $D$, as opposed simply to measuring how varying those
parameters changes $F(X)$. A thermodynamic integration method such as that employed in
Refs.~\onlinecite{matthews2012effect} and \onlinecite{poier2014influence} is well suited
for such a measurement. Finally, the effects of channel cross-section shape on the knot
behavior would be of interest to examine, as we have done previously in our study on backfolded
polymers under confinement in channels.\cite{polson2018free} We hope that experiments 
on knotted DNA will eventually be carried out to test the predictions of our simulations.

\begin{acknowledgments}
This work was supported by the Natural Sciences and Engineering Research Council of Canada (NSERC).  
We are grateful to Compute Canada and the Atlantic Computational Excellence Network (ACEnet) for 
use of their computational resources. We would like thank Alex Klotz for helpful discussions and
for a critical reading of the manuscript.
\end{acknowledgments}

\appendix
\section{Calculation of the free energy for conical confinement}
\label{app:a}

In principle, the multiple-histogram method used to calculate $F(X)$ for knotted polymers 
in cylinders in Sec.~\ref{subsec:cylin} can be employed to measure $F(Z_{\rm k})$, the 
knot-position dependence of the free energy for a polymer under confinement in a 
cone.  However, as noted in Sec.~\ref{sec:methods} previously, this approach suffers
from the presence of long correlation times associated with fluctuations in the polymer 
extension and, correspondingly, in the knot extension length, $X_{\rm k}$. Typically,
the correlation time is comparable to or greater than the entire simulation 
run time. Since the variation of $F$ with $Z_{\rm k}$ tends to depend
significantly on the knot extension, the contributions to $F$ from the
histogram associated with each window potential of Eq.~(\ref{eq:winpot})
are highly sensitive to the initial values of $X_{\rm k}$. These values
tend to be randomly distributed during the initialization routine of
the simulations, and so the resulting free energy functions tend to be of poor
quality. To circumvent this problem, we employ the multiple-histogram
method to measure $F(Z_{\rm k})$ for fixed $X$ (which essentially fixes the
knot span), and then carry out an appropriate average of these functions
for a number of values of $X$. The algorithm is described below.

Consider a knotted polymer confined to a cone aligned along the $z$ axis
with one end monomer tethered to $z$=0. Let $Z_{\rm k}$ and $X$ be the
knot position along $z$ and the polymer extension length, respectively. 
The probability distribution for the knot position, $\cP(Z_{\rm k})$, 
satisfies
\be
\cP(Z_{\rm k}) = \int \cP(Z_{\rm k}|X) \cP(X) dX,
\label{eq:PZk}
\ee
where $\cP(X)$ is the probability distribution for the polymer extension,
and where $\cP(Z_{\rm k}|X)$ is the conditional probability for the
knot position for a given extension length $X$. Here, the integral
is over all accessible values of $X$. Each probability distribution 
is related to a corresponding free energy function; that is,
\be
\cP(X) & = & \frac{\exp\left(-\beta F(X)\right)}{\int dX\,\exp\left(-\beta F(X)\right)}, 
\label{eq:P1}\\
\cP(Z_{\rm k}) & = & 
\frac{\exp\left(-\beta F(Z_{\rm k})\right)}{\int dZ_{\rm k}\exp\left(-\beta F(Z_{\rm k})\right)},
\label{eq:P2}
\ee
and
\be
\cP(Z_{\rm k}|X) & = & 
\frac{\exp\left(-\beta F(Z_{\rm k}|X)\right)}{\int dZ_{\rm k}\,\exp\left(-\beta F(Z_{\rm k}|X)\right)},
\label{eq:P3}
\ee
where $\beta\equiv 1/k_{\rm B}T$. It follows from Eqs.~(\ref{eq:PZk}) -- (\ref{eq:P3}) that: 
\be
\beta F(Z_{\rm k}) = -\ln\left[ \int dX\,C(X) \exp\left(-\beta (F(Z_{\rm k}|X)+F(X))\right)
\right], \nn \\
\label{eq:FZk}
\ee
where
\be
C(X) \equiv \left[\int \exp(-\beta F(Z_{\rm k}|X))\,dZ_{\rm k} \right]^{-1}.
\label{eq:CX}
\ee

We use Eq.~(\ref{eq:FZk}) to calculate the dependence of the free energy on
knot position in the simulations. To do so, the free energy function $F(X)$ is
calculated for a polymer in a cone using the same method as that employed for
cylindrical channels in Section~\ref{subsec:cylin}. The free energy function
$F(Z_{\rm k}|X)$ is calculated by constraining the extension length to a
particular value $X$ and then employing the multiple-histogram method described
in Section~\ref{sec:methods} to calculate the probability distribution for 
the knot position, $Z_{\rm k}$. The integrals of Eqs.~(\ref{eq:FZk}) and (\ref{eq:CX})
are approximated with discrete summations; thus,
\be
\beta F(Z_{\rm k}) \approx -\ln\left[ \sum_i C(X_i) 
\exp\left(-\beta (F(Z_{\rm k}|X_i)+F(X_i))\right) \right], \nn \\
\label{eq:FZk2}
\ee
where
\be
C(X_i) \equiv \left[\sum_{Z_{\rm k}} \exp(-\beta F(Z_{\rm k}|X_i))\right]^{-1}.
\label{eq:CX2}
\ee
The values of $X_i$ are chosen to lie between bounds defined such that 
$F(X_i)-F_{\rm min}<7k_{\rm B}T$, where $F_{\rm min}$ is the minimum of the
free energy function. The probability that $X$ lies outside this range is
negligible. Typically we choose 10--15 values of $X_i$ within this range.

%
% In addition to $F(Z_{\rm k})$, we want to know the dependence of the mean
% knot span with knot position, $\bar{X}_{\rm k}(Z_{\rm k})$. This can be approximated
% \be
% \bar{X}_{\rm k}(Z_{\rm k}) = \sum_{i} X_{\rm k}(Z_{\rm k},X_i) \cP(X_i|Z_{\rm k})
% \label{eq:XZ}
% \ee
% where $X_{\rm k}(Z_{\rm k},X_i)$ is the mean knot span for a given knot position $Z_{\rm k}$
% and a given polymer extension length $X_i$. As above, the summation is a discrete
% approximation for an integral of $X$ over all accessible values. The quantity
% $\cP(X_i|Z_{\rm k})$ is the conditional probability that the polymer has extension
% $X_i$ given that the knot extension is $Z_{\rm k}$. This quantity can be related
% to known quantities using the following basic property of probability distributions:
% \be
% \cP(X_i|Z_{\rm k})\cP(Z_{\rm k}) = \cP(Z_{\rm k}|X_i) \cP(X_i)
% \label{eq:Pbas}
% \ee
% Using Eqs.~(\ref{eq:P1}) -- (\ref{eq:P3}), (\ref{eq:XZ}), and Eq.~(\ref{eq:Pbas}), it is 
% easily shown that
% \be
% \bar{X}(Z_{\rm k}) = \frac{\sum_i X_{\rm k}(Z_{\rm k},X_i) e^{-\beta(F(Z_{\rm k}|X_i)+F(X_i))}}
% {\sum_i e^{-\beta(F(Z_{\rm k}|X_i)+F(X_i))}}.
% \ee
%

\section{Confinement free energy of a polymer in cone in the Odijk regime}
\label{app:b}

A theoretical estimate for the variation of the free energy of a knotted polymer with respect
to knot position is provided in Appendix~\ref{app:c}. The theoretical model used requires
the confinement free energy of an unknotted polymer under conical confinement, which we 
derive in this appendix.

Consider a polymer confined to a conical channel of half-angle $\alpha$ aligned along the 
$z$-axis. One end monomer is fixed at $z$=0, where the cone diameter is $D_0$. For $z\geq 0$,
the diameter is
\be
D(z) = D_0 + 2z\tan\alpha.
\label{eq:Dz}
\ee
We consider only channels with sufficiently small $D_0$ and $\alpha$ such that $D(z)\ll P$ for 
all locations where the monomers are present; that is, Odijk conditions are assumed to apply
for the entire span of the polymer along the channel.

For a polymer in a cylindrical tube with $\alpha$=0 and diameter $D_0$, the extension $X$ of 
a polymer of contour length $L$ satisfies
\bes
L = X/(1-b(D_0/P)^{2/3}),
\ees
where $b=0.17$.\cite{dai2016polymer} For the case of $\alpha>0$, we note that an infinitesimal
portion of the polymer of contour length $dL$ located at position $z$ with an extension $dz$ satisfies
\be
dL = dz/(1-b(D(z)/P)^{2/3}),
\label{eq:dLdz}
\ee
where $D(z)$ is given by Eq.~(\ref{eq:Dz}). It follows that
\bes
\int dL = L = \int_0^X \frac{dz}{1-b(D(z)/P)^{2/3}},
\ees
which yields the following relation between $L$ and $X$:
\bes
L = f(X),
\ees
where
\be
f(X) \equiv &&\left(\frac{3P}{2b^{3/2}\tan\alpha}\right) 
\left[ \tanh^{-1}\left(b^{1/2}\left(\frac{D(X)}{P}\right)^{1/3}\right) \right. \nonumber\\
&&- b^{1/2}\left(\frac{D(X)}{P}\right)^{1/3} 
-\tanh^{-1}\left(b^{1/2}\left(\frac{D_0}{P}\right)^{1/3}\right)  \nonumber\\
&& \left. + b^{1/2}\left(\frac{D_0}{P}\right)^{1/3} \right].
\label{eq:Lext}
\ee

In the Odijk regime the confinement free energy of a semiflexible polymer in a cylindrical
($\alpha$=0) channel of diameter $D_0$ is 
\be
F_{\rm c} = a L D_0^{-2/3}P^{-1/3},
\label{eq:Fod}
\ee
where $a$=2.3565.\cite{dai2016polymer} In the case of a conical channel, a small portion of 
the polymer contour length $dL$ at position $z$ contributes
\be
dF_{\rm c} = a dL (D(z))^{-2/3} P^{-1/3}.
\label{eq:dFc}
\ee
From Eqs.~(\ref{eq:dLdz}) and (\ref{eq:dFc}) it follows
\bes
dF_{\rm c} = \frac{a dz}{[(D(z))^{2/3} P^{1/3}] (1-b (D(z)/P)^{2/3})}.
\ees
Integration of along $z$ from $z$=0 to $z$=X gives the total free energy:
\be
F_{\rm c} = &&\left(\frac{3a}{2b^{1/2}\tan\alpha}\right)
      \left[\tanh^{-1}\left(b^{1/2}\left(\frac{D(X)}{P}\right)^{1/3}\right) \right.\nonumber\\
    && \left. - \tanh^{-1}\left(b^{1/2}\left(\frac{D_0}{P}\right)^{1/3}\right)\right],
\label{eq:Fcone}
\ee
where the extension length $L$ is determined by Eq.~(\ref{eq:Lext}).

\section{Theoretical model for the free energy function of a knot in a cone}
\label{app:c}

\begin{figure}[!ht]
\begin{center}
\vspace*{0.2in}
\includegraphics[width=0.35\textwidth]{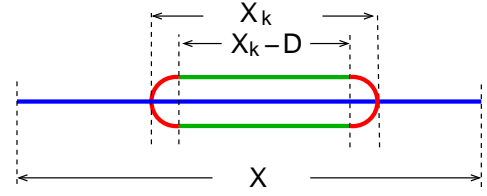}
\end{center}
\caption{Illustration of the model used for theoretical prediction for a knot in a cone. The
blue line represents an unknotted linear polymer of extension length $X$. The linear polymer 
overlaps with a ring polymer of extension $X_{\rm k}$. The diameter of the ring polymer 
hairpins (colored red) is approximately $D$, the diameter of the cone at the center of the
the ring. The extension of the extended sections of the ring polymer (colored green) is
thus $X_{\rm k}-D$.}
\label{fig:loopmodel}
\end{figure}

In this appendix we derive an expression for the variation of the free energy with respect 
to knot position of a knotted polymer under conical confinement.  To do so, we model the 
knotted polymer as an unknotted linear polymer of span $X$ overlapping a ring polymer of 
span $X_{\rm k}$, as illustrated in Fig.~\ref{fig:loopmodel}. The ring polymer is an 
approximation for the knot in the real system, each of which has two hairpin turns. The
diameter of the hairpin is chosen to be the local diameter of the cone, $D$. As noted
in Sec.~\ref{subsec:cylin} this approximation is likely only slightly overestimate under
Odijk conditions.\cite{odijk2006dna,chen2017conformational} The combined contour lengths 
of the linear and ring polymers are chosen to be equal to that of the real polymer.

We identify four main contributions to the free energy:
\begin{enumerate}
\item
The free energy of the two hairpins, $F_1$.
\item
The overlap free energy of the three subchains that lie between the two hairpins
along the channel, $F_2$.
\item
The confinement free energy of the two extended sections of the ring polymer (colored green
in the figure), $F_3$.
\item
The confinement free energy of the linear polymer (colored blue in the figure), $F_4$.
\end{enumerate}

For the hairpin free energy, we use the results of a study by Chen.\cite{chen2017conformational}
In that study, a numerical solution to the Green's function equations for an ideal chain confined 
to a channel yielded a hairpin free energy that could be approximated with the following equation:
\be
F_{\rm hp} = \frac{2E_m}{\tilde{D}} -
\frac{3}{2}\ln\left[\frac{A_2\tilde{D}+A_0\tilde{D}^2}{1+A_1\tilde{D}+A_0\tilde{D}^2 }\right] + \ln 4,
\ee
where $\tilde{D}\equiv D/P$ and where the dimensionless numerical factors are $E_m=1.43557$, 
$A_0=1.0410$, $A_1=-0.6046$ and $A_2=1.2150$.  For simplicity, we neglect the small variation 
of the $D$ over the span of the knot, which is located at position $Z_{\rm k}$. 
(Note that this approximation is valid only for very small cone angles. Carrying out
calculations with and without it produced results with negligible difference for the
cone angles used here.) Thus, the hairpin free energy is:
\be
F_1(Z_{\rm k}) = 2F_{\rm hp}(D(Z_{\rm k})),
\label{eq:F1}
\ee
where the factor of 2 accounts for fact that there are two hairpins, and where
\be
D(Z_{\rm k}) = D_0 + 2Z_{\rm k}\tan\alpha.
\label{eq:Dz2}
\ee

As noted in Sec.~\ref{subsec:cylin}, the overlap free energy for an S-loop or knot
for a polymer confined to a cylinder in the Odijk regime is approximately
\bes
F_{\rm ov} = C D^{-5/3} P^{1/3} (X_{\rm k} - D),
\ees
where $X_{\rm k}-D$ is the span of the three overlapping polymer strands in the knot
excluding the hairpins, and where the constant is estimated to be $C$=9.45. Thus,
\be
F_2(Z_{\rm k},X_{\rm k}) = C (D(Z_{\rm k}))^{-5/3} P^{-1/3} (X_{\rm k} - D(Z_{\rm k})).~~~~
\label{eq:F2}
\ee

For the contribution from the confinement free energy of the two extended portions of the ring, 
we neglect the small variation of the cone diameter along the span of the knot. The overlap free 
energy is thus
\be
F_3(Z_{\rm k},X_{\rm k}) = 2a (X_{\rm k}-D(Z_{\rm k})) \left(D(Z_{\rm k})\right)^{-2/3} P^{-1/3}, ~~~~
\label{eq:F3}
\ee
where $a$=2.3565.\cite{dai2016polymer} In addition, the factor of 2 is due to the presence
of two extended strands of the ring polymer, and $D(Z_{\rm k})$ is given by Eq.~(\ref{eq:Dz2}).

Finally, consider the free energy of the linear polymer in the cone. Since the contour
length of the knotted polymer $L$ is the sum of the contour length for the linear
polymer, $L^\prime$, and that of the ring polymer, it follows that:
\bes
L^\prime = L - \pi D(Z_{\rm k}) - 2(X_{\rm k}-D(Z_{\rm k})).
\ees
Thus,
\be
L - (\pi-2)D(Z_{\rm k}) - 2X_{\rm k}  = f(X),
\ee
where $f(X)$ is given by Eq.~(\ref{eq:Lext}).  It follows that:
\be
X(Z_{\rm k},X_{\rm k}) = f^{-1}(L-(\pi-2)D(Z_{\rm k}) - 2X_{\rm k}).
\label{eq:Xfm1}
\ee
Using Eq.~(\ref{eq:Fcone}), the confinement free energy of the linear polymer in the cone is thus:
\be
F_4(Z_{\rm k},X_{\rm k}) & = & \left(\frac{3a}{2b^{1/2}\tan\alpha}\right)
      \left[\tanh^{-1}\left(b^{1/2}\left(\frac{D(X)}{P}\right)^{1/3}\right)\right.
\nonumber \\
    && \left. - \tanh^{-1}\left(b^{1/2}\left(\frac{D_0}{P}\right)^{1/3}\right)\right],
\label{eq:F4}
\ee
where the dependence of $F_4$ on $Z_{\rm k}$ and $X_{\rm k}$ arises from the relation for the extension 
length $X$ in Eq.~(\ref{eq:Xfm1}).

\begin{figure}[!ht]
\begin{center}
\vspace*{0.2in}
\includegraphics[width=0.4\textwidth]{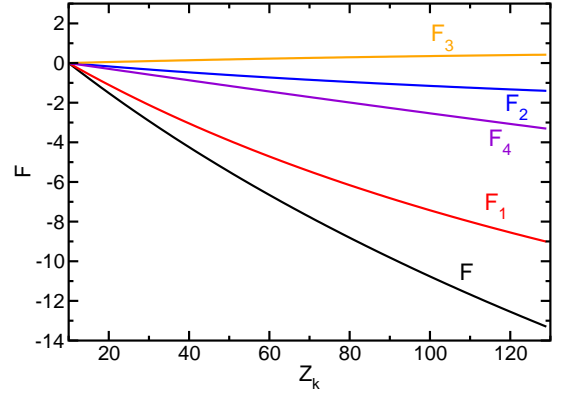}
\end{center}
\caption{Variation of each contribution to the predicted free energy with knot position.
The contributions $F_1$, $F_2$, $F_3$ and $F_4$ are calculated using Eqs.~(\ref{eq:F1}),
(\ref{eq:F2}), (\ref{eq:F3}), and (\ref{eq:F4}), respectively, and $F$ is the sum
of the contributions. The predictions are calculated for a $3_1$ knot, $N$=200, $\kappa=15$, 
$D_0$=4.0, and $\alpha$=$0.57^\circ$.  }
\label{fig:Fth_all}
\end{figure}

The total free energy if the knotted polymer is given by the sum
\be
F(Z_{\rm k}) = F_1(Z_{\rm k}) + F_2(Z_{\rm k},X_{\rm k}) + F_3(Z_{\rm k},X_{\rm k}) 
+ F_4(Z_{\rm k},X_{\rm k}),
\nonumber \\
\ee
where the free energy contributions are given by Eqs.~(\ref{eq:F1}), (\ref{eq:F2}), (\ref{eq:F3})
and (\ref{eq:F4}). Finally, the dependence of $F$ on the knot span $X_{\rm k}$ must be removed.
Note that $X_{\rm k}$ is a fluctuating variable whose mean and variance depends on $Z_{\rm k}$, 
the position of the knot along the channel. To estimate the variation of $X_{\rm k}$ with 
$Z_{\rm k}$, we have chosen the following procedure. A set of simulations for a knotted polymer 
in a cylindrical channel were carried out to measure $F(X)$ and $X_{\rm k}(X)$ for various values 
of channel diameter $D$. At each $D$, the mean value of $X_{\rm k}$ was calculated
\be
\bar{X}_{\rm k}(D) = \frac{\int X_{\rm k}(X;D) e^{-\beta F(X;D)} dX}{\int e^{-\beta F(X;D)} dX},
\ee
where the integrals were approximated using discrete summations. Applying  this result to
the conical channel requires the $Z_{\rm k}$-dependence of $D$, which is provided by Eq.~(\ref{eq:Dz2}).
Figure~\ref{fig:Fth_all} shows a comparison of each of the contributions for a $3_1$ knot for
a system with $N$=200, $\kappa=15$, $D_0$=4.0, and $\alpha$=$0.57^\circ$. The hairpin contribution
to free energy is the dominant term, a consequence of the narrowness of the conical channel.

% \bibliography{paper}
% \bibliographystyle{apsrev4-1}


\end{document}

\title{Supplemental Material for: ``Free Energy of a Knotted Polymer Confined to Narrow Cylindrical 
and Conical Channels''}

\author{James M. Polson}
%\email{jpolson@upei.ca}
\affiliation{Department of Physics, University of Prince Edward Island,
550 University Ave., Charlottetown, Prince Edward Island, C1A 4P3, Canada}
\author{Cameron Hastie}
\altaffiliation{Current address: Department of Physics, McGill University, 3600 rue University,
Montreal, Quebec, H3A 2T8, Canada}
\affiliation{Department of Physics, University of Prince Edward Island,
550 University Ave., Charlottetown, Prince Edward Island, C1A 4P3, Canada}

%\date{\today}

\maketitle

In Section~IV A of the article, we examined the variation of the mean knot extension
length $X_{\rm k}$ with polymer extension length $X$. Figure~7(a) shows results for a 
polymer of length $N$=400, bending rigidity $\kappa$=15, and cylinder diameters ranging
from $D$=4 to $D$=7.  In the region of $X$ where the free energy $F$ increases linearly 
with decreasing $X$, we find that ${X}_{\rm k}$ also increases linearly with decreasing 
$X$. Here, $dF/dX \approx -0.5$ for all $D$, and ${X}_{\rm k}$ decreases monotonically
with $D$ at any given $X$. To account for the variation with $D$, we presented a simple
scaling argument that predicted the quantity
\be
\zeta_{\rm k} & \equiv & {X}_{\rm k} + ({\textstyle \frac{\pi}{2}}-1)D
            + {\textstyle\frac{1}{2}} \alpha_\parallel L D^{2/3} P^{-2/3}
\ee
scales as $\zeta_{\rm k}=(L-X)/2$, independent of $D$ and $P$. This expectation was borne 
out by the data collapse of Fig.~7(b). The scaling argument neglected two potentially
important system properties: (1) fluctuations in the extension length of the polymer, 
and (2) interactions between elongated parts of the polymer that overlap in the knot or 
S-loop. The purpose of this Supplemental Material document is to provide a 
more rigorous justification of the prediction and show that the effects of these two system 
properties are negligible.

As in the article, we use established results for polymer confinement in the
Odijk regime, where $D\ll P$.  Note that the required condition
$D\ll P$ is only marginally satisfied in this case ($P/D = 2.14-3.75$), and thus
some quantitative discrepancy between the predicted and observed behavior is to be
expected. Since the results for knotted ($3_1$) and unknotted (S-loop) polymers
are identical over the regime of interest ($X<X_{\rm min}$), we ignore the effects
of topology. 

First, consider a polymer with no backfolding. Recall that the mean extension length 
of a polymer in this regime is given by
\be
\bar{L}_{\parallel}=L(1-\alpha_{\parallel} D^{2/3}P^{-2/3}),
\ee
where $L$ is the contour length of the polymer and where the prefactor is
$\alpha_{\parallel}=0.1701\pm 0.0001$.\cite{dai2016polymer} In addition, the
variance in the extension is approximately $\sigma_L^2\approx \alpha_\delta LD^2P^{-1}$,
where the prefactor for cylindrical channels is $\alpha_\delta=0.0150\pm 0.0002$.%
\cite{dai2016polymer} Thus, the free energy function for arbitrary $L_{\parallel}$ can
be written as 
\be
F(L)=\frac{1}{2}k_{\rm sp} (L_\parallel-\bar{L}_{\parallel})^2,
\label{eq:FL}
\ee
where the spring constant is given by $k_{\rm B}\alpha_\delta^{-1} TL^{-1}D^{-2}P$.\cite{dai2016polymer}

\begin{figure}[!ht]
\begin{center}
\vspace*{0.2in}
\includegraphics[width=0.6\textwidth]{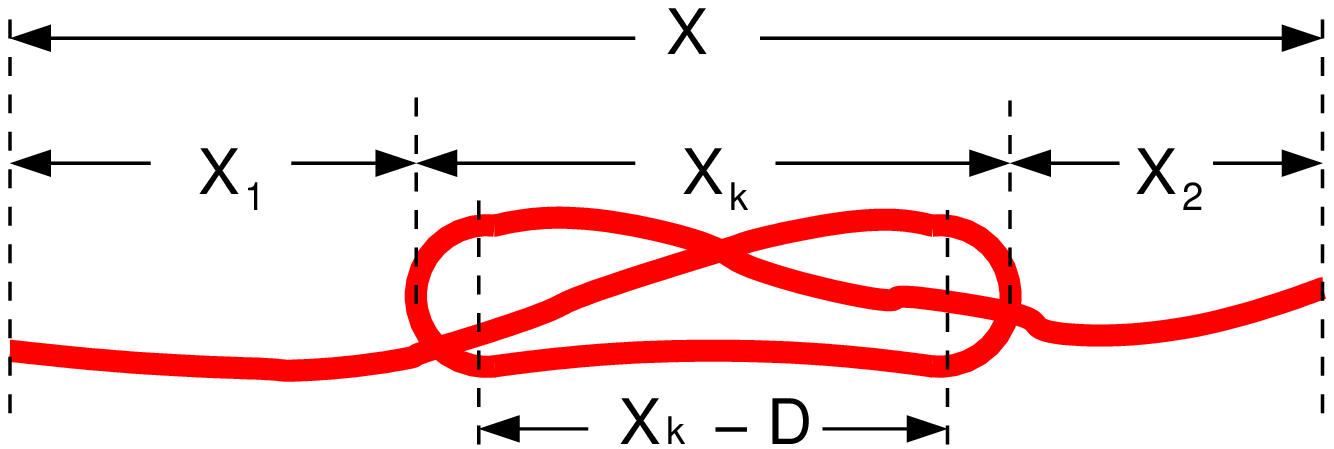}
\end{center}
\caption{Illustration of the quantities $X_1$, $X_2$, $X_{\rm k}$ and $X$ described
in the text.  }
\label{fig:Xillust}
\end{figure}

Now, consider a polymer with two hairpin folds, which may result from an S-loop or a knot. We 
first define the effective contour length as $\ell\equiv L-\pi D$. This is the contour length
of all elongated portions of the polymer, i.e., excluding the contour in the 
two hairpins.  Here, we assume the hairpin diameter is $D$, which is likely only a slight 
overestimate in the narrow-channel limit.\cite{odijk2006dna,chen2017conformational}
The span of the S-loop/knot, $X_{\rm k}$, is the mean distance between the two hairpins.
Defining $X_1$ and $X_2$ as the distance of each hairpin to the nearest polymer end,
the effective extension of the polymer can be defined as 
\bes
\ell_\parallel\equiv X_1 + X_2 + 3X_{\rm k}-2D.
\ees
The effective extension is the sum of the extension lengths
of all individual elongated portions of the polymer outside the hairpins.
Since $X=X_1+X_2+X_{\rm k}$, it follows that 
\be
\ell_\parallel = 2X_{\rm k} + X - 2D.
\ee
Figure~\ref{fig:Xillust} provides an illustration of these various quantities.
The generalization of Eq.~(\ref{eq:FL}) is thus
\begin{eqnarray}
F_1(X_{\rm k};X) & = & {\textstyle{\frac{1}{2}}} k_{\rm sp}\left(\ell_\parallel
- \bar{\ell}_\parallel\right)^2
\nonumber \\
& = & 2k_{\rm B}T(\alpha_\delta)^{-1}\ell^{-1}D^{-2} P(X_{\rm k}+{\textstyle \frac{1}{2}}X
                       - D- {\textstyle \frac{1}{2}}\ell
      + {\textstyle \frac{1}{2}}\alpha_\parallel \ell D^{2/3}P^{-2/3})^2.
\label{eq:F1X}
\end{eqnarray}

A second contribution to the free energy arises from the overlap in the three strands that
lie between the hairpins of the knot or S-loop. As noted previously,\cite{polson2017free,%
polson2019polymer} the approximate form of this free energy can be written
\begin{eqnarray}
F_2(X_{\rm k};X) & = & C D^{-5/3} P^{-1/3} (X_{\rm k} - D)
\label{eq:Fov}
\end{eqnarray}
where the scaling exponents are derived using a 2nd-virial approximation for interacting Odijk
deflection segments, which are modeled as hard cylinders. A further addition to the free energy
is that associated with the two hairpin turns. However, the hairpin properties are not expected
to be affected by variations in $X$, and thus this constant free energy contribution can be omitted. 
The total free energy is the sum of the two contributions given in Eqs~(\ref{eq:F1X}) and 
(\ref{eq:Fov}):
\begin{eqnarray}
F(X_{\rm k};X) = F_1(X_{\rm k};X) + F_2(X_{\rm k};X).
\end{eqnarray}
Minimizing $F$ by setting $dF/dX_{\rm k}=0$, we find a predicted knot extension length of
\bes
{X}_{\rm k} & \approx & -{\textstyle\frac{1}{2}} X + {\textstyle\frac{1}{2}} L
            - ({\textstyle \frac{\pi}{2}}-1)D
            - {\textstyle\frac{1}{2}} \alpha_\parallel (L-\pi D) D^{2/3} P^{-2/3}
            - C \alpha_{\delta} (L-\pi D) D^{1/3} P^{-4/3}/4\pi k_{\rm B}T.
\ees
Since $L\gg \pi D$ for our simulations, this simplifies to
\be
{X}_{\rm k} & \approx & -{\textstyle\frac{1}{2}} X + {\textstyle\frac{1}{2}} L
            - ({\textstyle \frac{\pi}{2}}-1)D
            - {\textstyle\frac{1}{2}} \alpha_\parallel L D^{2/3} P^{-2/3}
            - C \alpha_{\delta} L D^{1/3} P^{-4/3}/4\pi k_{\rm B}T.
\label{eq:Xk}
\end{eqnarray}

The first term accounts for the observed
slope of $dX_{\rm k}/dX\approx -0.5$, while the third, fourth and fifth terms account for the
observed decrease in $X_{\rm k}$ with increasing $D$.  To determine the constant $C$, we use
results for the free energy function for an S-loop in Figs.~3 and 4 of the article.
The linear regime corresponds to a constant value of $m\equiv dF/dX$.
Since Eq.~(\ref{eq:Xk}) implies $dX_{\rm k}/dX=-\frac{1}{2}$, it follows that
$m = -\frac{1}{2}CD^{-5/3}P^{-1/3}$ and so $C=-2mD^{5/3}P^{1/3}$. Using results for $P$=15
and $D$=4, where it was found that $m=-0.19$, we estimate that $C=9.45$. With this value of
the constant, it is easily verified that the last term in Eq.~(\ref{eq:Xk}) is negligible
compared to the others. Essentially, this implies that the effects of fluctuations 
of the knot extension length  in the Odijk regime are negligible, as are the
interactions between overlapping elongated subchains inside the knot. Defining the shifted
knot extension, $\zeta_{\rm k}$, as
\begin{eqnarray}
\zeta_{\rm k} & \equiv & {X}_{\rm k} + ({\textstyle \frac{\pi}{2}}-1)D
            + {\textstyle\frac{1}{2}} \alpha_\parallel L D^{2/3} P^{-2/3},
\label{eq:Xkdag}
\end{eqnarray}
it follows from Eq.~(\ref{eq:Xk}) and from discarding the negligible final term in that equation
that $\zeta_{\rm k}=(L-X)/2$ for all values of $D$, $P$, independent of the  polymer extension
$X$. 

% \bibliography{paper}
% \bibliographystyle{apsrev4-1}
